IAC-22,B4,2,9,x69501

# The Star-Planet Activity Research CubeSat (SPARCS): Determining Inputs to Planetary Habitability

**David R. Ardila*[a], Evgenya Shkolnik[b], Paul Scowen[c], Daniel Jacobs[b], Dawn Gregory[d], Travis Barman[e], Christopher Basset[a], Judd Bowman[b], Samuel Cheng[a], Jonathan Gamaut[b], Logan Jensen[b], April Jewell[a], Mary Knapp[f], Matthew Kolopanis[b], Joseph Llama[g], R. O. Parke Loyd[h], Victoria Meadows[i], Shouleh Nikzad[a], Sara Peacock[c], Tahina Ramiaramanantsoa[b], Nathaniel Struebel[d], Mark Swain[a].**

[a]*Jet Propulsion Laboratory, California Institute of Technology, Pasadena, CA, USA. david.r.ardila@jpl.nasa.gov.*
[b]*Arizona State University, Tempe, AZ, USA*
[c]*NASA Goddard Space Flight Center, Greenbelt, MD. USA*
[d]*AZ Space Tech, Tempe, AZ, USA*
[e]*University of Arizona, Tucson, AZ, USA*
[f]*Massachussetts Institute of Technology, Haystack Observatory, Westford, MA, USA*
[g]*Lowell Observatory, Flagstaff, AZ, USA*
[h]*Eureka Scientific Inc. Oakland, CA, USA*
[i]*University of Washington, Seattle, WA, USA*

\* Corresponding Author

**Abstract**

Seventy-five billion low-mass stars in our galaxy host at least one small planet in their habitable zone (HZ). The stellar ultraviolet (UV) radiation received by the planets is strong and highly variable, and has consequences for atmospheric loss, composition, and habitability. These effects are amplified by the extreme proximity of the stellar HZs (0.1- 0.4 AU) in low-mass stars.

SPARCS is a NASA-funded mission to characterize the quiescent and flare UV emission from low-mass stars. SPARCS will observe 10 to 20 low-mass stars, over timescales of days, simultaneously in two UV bands: 153-171 nm and 260-300 nm. SPARCS Sun-synchronous terminator orbit allows for long periods of uninterrupted observations, reaching 10s of days for some targets. The payload consists of a 10 cm-class telescope, a dichroic element, UV detectors and associated electronics, a thermal control system, and an on-board processor. The payload is hosted on a Blue Canyon Technologies 6U CubeSat.

SPARCS hosts several technology innovations that have broad applicability to other missions. The payload demonstrates the use of "2D-doped" (i.e., delta- and superlattice-doped) detectors and detector-integrated metal dielectric filters in space. This detector technology provides ~5x larger quantum efficiency than NASA's GALEX detectors. In addition, SPARCS' payload processor provides dynamic exposure control, automatically adjusting the exposure time to avoid flare saturation and to time-resolve the strongest stellar flares. A simple passive cooling system maintains the detector temperature under 238K to minimize dark current. The spacecraft bus provides pointing jitter smaller than 6", minimizing the impact of flat-field errors, dark current, and read-noise. All these elements enable competitive astrophysics science within a CubeSat platform.

SPARCS is currently in the final design and fabrication phase (Phase C in the NASA context). It will be launched in 2024, for a primary science mission of one year.

**Keywords:** Ultraviolet, Astrophysics, CubeSats, NASA, Exoplanets, Stars

**Acronyms/Abbreviations**
AIMO: Advanced Inverted Mode Operation
APRA: Astrophysics Research and Analysis
BCT: Blue Canyon Technologies
FUV: Far Ultraviolet
GALEX: Galaxy Evolution Explorer
HST: Hubble Space Telescope
HZ: Habitable Zone
JPL: Jet Propulsion Laboratory
QE: Quantum Efficiency
LEO: Low-Earth Orbit
NUV: Near Ultraviolet
SAA: South Atlantic Anomaly
SPARCS: Star-Planet Activity Research CubeSat
TRL: Technology Readiness Level
TEC: Thermoelectric Cooler
UV: Ultraviolet

## 1. Introduction

Low-mass stars, with masses between 60% and 8% that of the Sun, are the most common type of stars in the Galaxy. They are also known as M-dwarfs, or Red Dwarfs. Based on statistics of planet abundance, astronomers believe that there are 75 billion low-mass stars in the Galaxy, hosting at least one planet like the Earth in the Habitable Zone (HZ [1]).

The stellar ultraviolet (UV) radiation from these stars, incident upon the planets, is strong and highly variable, and impacts planetary atmospheric loss, composition, and habitability (Figure 1). These effects are amplified by the extreme proximity of their HZs (0.1–0.4 AU; [2]).

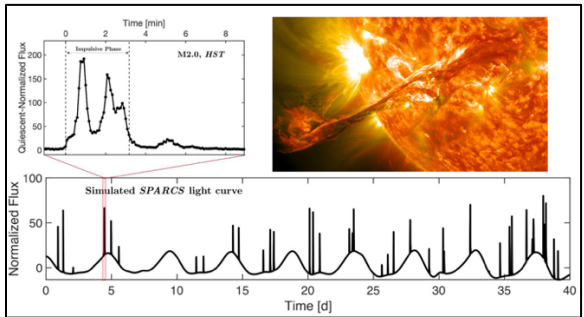

Figure 1. *Top-Left:* M-type stars flare often, and their flares are quite strong. The figure shows an actual super flare observed with the Hubble Space Telescope (HST) [3]. *Top-Right:* Flaring takes place in all kinds of stars. Actual flare observed in the Sun, from the Solar Dynamics Observatory (SDO). Image credit: NASA/GSFC/SDO. *Bottom:* Simulated light curve from a M-type star. The strong flares are superimposed on the periodic rotational variability.

The UV flux emitted during the stellar lifetime drives water loss and photochemical O2 build-up for terrestrial planets within the HZ. [4, 5]. UV-driven photochemistry strongly affects a planet's atmosphere (e.g., [6]), could limit the planet's potential for habitability, and may confuse studies of habitability by creating false chemical biosignatures.

Our proposed observatory, the Star-Planet Activity Research Cube-Sat (SPARCS), will be the first mission to provide the time-dependent spectral slope, intensity, and evolution information of M-star UV radiation. Extending UV time-domain knowledge of the stars, from the timescales of tens of hours (the current limit) to a timescale of months to a year is crucial to characterize the stellar behaviour. A dedicated monitoring experiment, such as SPARCS, is the only way achieve this.

SPARCS will also demonstrate the long-term performance in space of JPL-developed 2D-doped process [7], which dramatically increases the Quantum Efficiency (QE) of silicon detectors in the UV. Environmental testing on the ground has shown that instrument systems with 2D-doped detectors are at Technology Readiness Level (TRL [8]) 6, the standard needed for infusion into a science mission.

## 2. The SPARCS Payload

The SPARCS payload block diagram and implementation are shown in figure 2. Payload parameters are in Table 1.

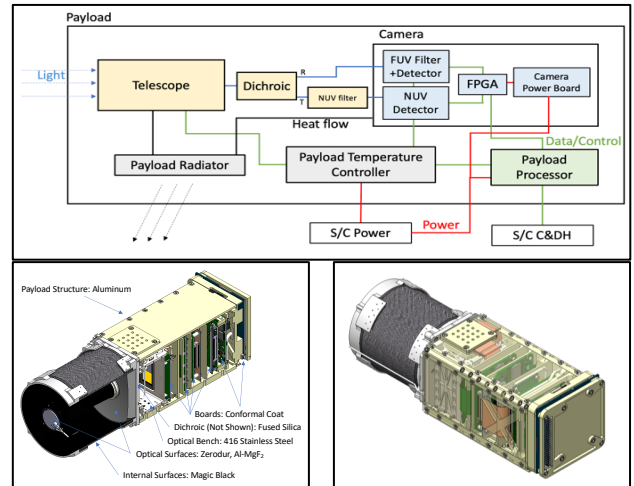

Figure 2. *Top:* Block diagram for the SPARCS Payload. The payload processor implements automatic exposure control, while the payload temperature controller manages the thermal system. *Bottom:* The SPARCS payload is very compact, and fits within 3U of the spacecraft. It has a dedicated radiator (not shown), isolated from the spacecraft. Materials have been chosen for their low outgassing characteristics, to prevent optical contamination. Venting paths allow payload outgassing to space.

Table 1. Instrument Parameters

| Telescope | | |
|---|---|---|
| Effective Aperture | 8.6 cm | Inc. spider obscuration |
| F/ | 6 | |
| FOV | 40' diameter | |
| PSF | 20" | At field center. Includes jitter and drift (10 min.) |
| Detectors | | |
| Type | 2 x Teledyne-e2v CCD 47-20 | 2D-doped process results in 100% internal QE |
| Size | 1056 x 2069 pix. | Half is used for frame transfer |
| Pitch | 13 µm | 5.6"/pix |
| FUV Channel | | |
| Wave. at Max throughput | 157 nm | Filter is applied to the detector |
| FWHM | 24 nm | |
| S/N=10 in 10 min. | 1200 µJy | |
| NUV Channel | | |
| Wave. at Max throughput | 283 nm | Filter is applied to the dichroic |
| FWHM | 49 nm | |
| S/N=10 in 10 min. | 1000 µJy | |

73rd International Astronautical Congress (IAC), Paris, France, 18-22 September 2022.

*2.1. Telescope and aft optic*

The optical train is shown in figure 3. The telescope (Figure 4) is a Ritchey-Cretien design, with a primary aperture of 9.3 cm in diameter. The mirrors are Zerodur coated with Al-MgF$_2$. The telescope structure is composed of Invar 36, Titanium, and Carbon fibre. The telescope was fabricated by SigmaSpace, a subsidiary of Hexagon Federal.

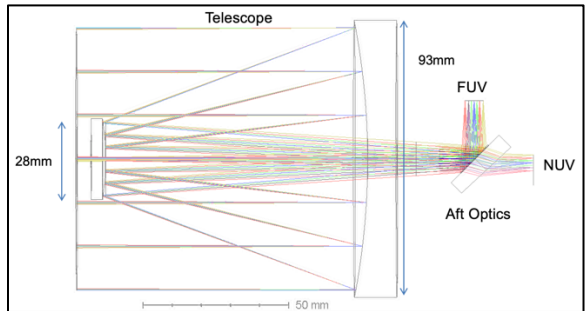

Figure 3. SPARCS Optical Train. The Aft optics element hosts the dichroic, the NUV filter, and performs aberration correction.

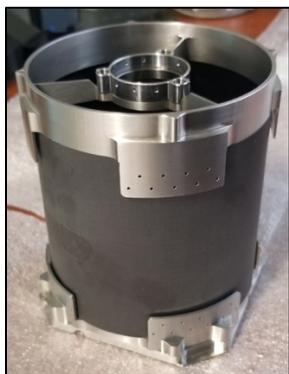

Figure 4. The SPARCS telescope is RC design, with a carbon fibre tube coated in anti-reflection paint.

After passing through the telescope, the light encounters a 2.5 cm diameter piece of UV-grade fused silica. The front surface of this aft optic is coated with a dichroic filter. The reflected beam is directed to the FUV detector and the transmitted beam to the NUV detector. The back surface is coated with the NUV filter. The two surfaces of the aft optic are at a 1.95 deg. angle, to partially correct for astigmatism in the transmitted beam. The aft optic and the dichroic coating were fabricated and applied by Teledyne-Acton Optics and Coatings. The NUV filter was applied to the aft optic by Materion Precision Optics (Figure 5).

At the centre of the 40' diameter field, the optical system is expected to result in Point Spread Functions (PSFs) Full Width at Half Maximum (FWHM) of 2.5" (FUV) and 6.2" (NUV). When including spacecraft jitter and drift the PSF FWHM is expected to be 20" (3.6 pix) for both bands (Section 3).

*2.2. FUV Filter and Detectors*

SPARCS' uses two Teledyne e2v CCD47-20 detectors. These are high performance, advanced inverted mode operation (AIMO), back illuminated, frame transfer detectors, with 100% filling factor. Frame transfer allows an image to be read slowly, thereby reducing read-noise, while the next image is being integrated.

The detectors are 2D-doped [7]. This JPL-developed process inserts a layer of Boron atoms on the back of the detector to eliminate the surface potential well. This results in detectors with 100% internal QE. The total QE is limited by the performance of the anti-reflection coating or filter, applied to the detector.

The FUV filter is a Fabry-Perot filter, made up of metal (Al) - dielectric (AlF$_3$) layers, directly applied to the detector using an Atomic Layer Deposition (ALD) process. The result is a FUV filter-detector system with a peak QE close 40%. This is to be compared to the 10%-20% detector QE from the Galaxy Evolution Explorer (GALEX), NASA's UV mission most like SPARCS (Figure 5).

The camera system is described in more detail in [9].

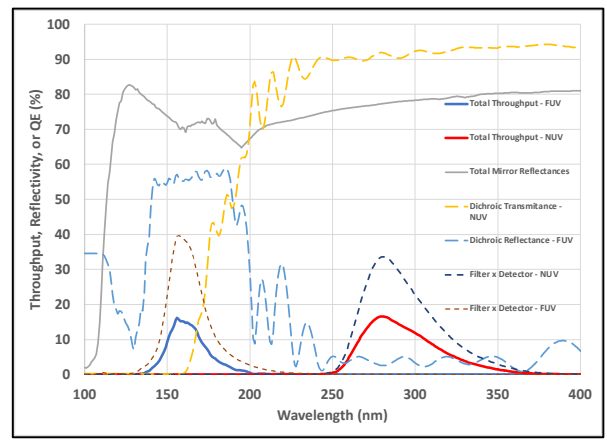

Figure 5. Component and system reflectivities, throughputs, or QE. SPARCS maximum throughput is ~15% in both bands, compared with GALEX's 3% maximum NUV throughput [10].

*2.3. On-board processing and Thermal Control*

In addition to the electronics associated with the detector, SPARCS carries an on-board processor and a payload temperature controller.

As shown in Figure 1, flares can result in brightness changes that are several hundred times the value of the quiescent stellar flux. To avoid saturating the detector, the on-board processor analyses the stellar photometry to determine if a flare is taking place. If so, it adjusts the exposure time and the detector gain for the next exposure, to avoid saturation. SPARCS is the first CubeSat mission to implement this type of dynamical exposure analysis [11].

The processor is a BeagleBone Black (BBB) with a Motherboard Module 2, manufactured by Pumpkin Space Systems. The BBB has a 1-GHz processor with 512 MB RAM. The processor runs an embedded Linux system generated with Buildroot, while the payload is controlled using custom Rust software based on prototyping in Python [11].

The thermal control system keeps the SPARCS detectors at 238±3K, to control detector dark current. Within the payload, the optical bench is isolated from the payload structure via G10 washers. The detector boards are isolated from the optical bench. A thermoelectric cooler (TEC), controlled by a Meerstetter TEC-1091 bi-polar DC controller, is used to cool the detectors during data collection and rejects the heat to the isolated payload radiator (530 cm$^2$), which faces cold space. The telescope, optical bench, and dichroic are passively cooled via the telescope.

The thermal system also allows for annealing, a process in which the detectors, telescope mirrors, and aft optics are raised to 50 C, for 24 - 48 hrs, to restore bad pixels, keep radiation damage from increasing the dark current, and decrease surface contamination. During annealing, the TEC is reverse biased and, along with the heaters, used to heat up the detectors.

## 2.4. System Sensitivity

Table 1 lists the system sensitivity, expressed as the brightness of a point source that results in a signal-to-noise (S/N) of 10 in 10 minutes. The error budget includes the contribution of the detector dark current and read noises (the dominant sources of noise), and flat field errors. The noise budget also includes noise due to astrophysical sources such the star itself and the Low Earth Orbit (LEO) UV background due to the Earth's exosphere emission, earthshine, and zodiacal light.

Because SPARCS' targets emit most of their light in the infrared, noise due to "red leak" is also included in the error budget. While the total system throughput in the infrared is small (e.g., 1 x 10$^{-6}$ in the FUV channel at 1μm), between 10% and 20% of photons detected in the UV science bands will come from longer wavelengths, depending on the target. This "red leak" effect is a nuisance background that can be subtracted from the observations but contributes to the overall noise.

The instrument will be characterized on the ground, and its performance verified in space. Absolute calibration is provided by observations of a network of white dwarfs, as was done for GALEX.

## 3. SPARCS Spacecraft and Mission

The SPARCS payload will be integrated at ASU with the Blue Canyon Technologies (BCT) XB6, 6U CubeSat bus, incorporating the XB1 bus avionics (Figure 6).

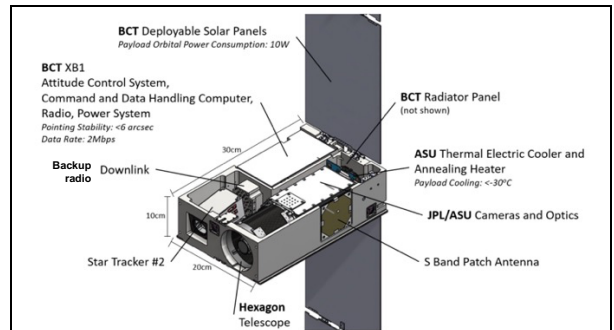

Figure 6. SPARCS payload integrated to the BCT SPACECRAFT. Communications provided via S-band. Ground stations are provided by KSAT-lite. Solar panels deploy in space. Two star trackers plus low-jitter reaction wheels provide pointing stability to <6".

In order to control read and dark current noise, SPARCS requires high frequency pointing errors of <6" (1σ). The XB1 avionics were selected because they have demonstrated this performance in an analogous system [12]. SPARCS will use dual star trackers, low jitter reaction wheels, torque rods, and integrated attitude control algorithms.

The system is not controlled for thermoelastic variability. While the thermal environment is benign, with eclipses taking place only once a year, we assume a drift of 1"/min. during integrations, based on the experience by [12].

To control contamination, spacecraft and payload will be integrated at ASU, on a dedicated facility built specially for SPARCS. Day-in-the-life as well as protoflight vibration and thermal testing will be performed as part of final integration. For more details see [13].

SPARCS will be launched in 2024, as a secondary payload, to a Sun-synchronous terminator orbit. After in-orbit checkout, SPARCS will commence its primary science mission of 1 year. Because there are no consumables on board, the mission may be extended if necessary.

The orbit allows for long observing periods for each target, ranging from about 60 min. to days, depending on the target location and time of year. For most of the year, the spacecraft will be fully illuminated by the Sun, mitigating the possibility of large thermoelastic variations. Observations will be interrupted when the target is on eclipse, for data downlink, momentum dumping, and for thermal management. The last one involves rotating the spacecraft 180 degrees every half orbit, to point the payload radiator to cold space. The spacecraft will be operated during passages of the South Atlantic Anomaly (SAA), but it is likely that data will not be usable in these passages. Overall, observation efficiency is expected to be >50%.

During the science mission, ~10 - 20 stars will be observed, continuously, over periods of time commensurate with their rotation periods (20 days average), to map their short (flares) and long (rotation)-term UV variability. Exposures range from 2 sec. to 10 min. After each exposure, the payload processor performs a quick data reduction to determine if the star is flaring and adjusts the exposure time accordingly [11].

Postage stamp data for each target and for other targets in the field, as well as occasional full-frame images, will be downlinked to the ground on two contacts per day. The data will be fully processed on the ground, archived in the Mikulski Archive for Space Telescopes (MAST), and made public after processing and validation.

## 4. Conclusions

SPARCS is a mission designed to understand the UV radiation output of stars at short and long timescales. This radiation can have significant consequences on planet habitability and on astronomers' ability to interpret planetary spectra when looking for biosignatures.

SPARCS will demonstrate in space the performance of 2D-doped detectors developed by JPL. This is a surface-treatment process that results in 100% internal QE and can be applied to any type of silicon detector.

The SPARCS payload consists of a 10 cm-class telescope, a dichroic, two UV detectors and associated electronics, a payload processor, and a thermal control processor, as well as associated thermal control and mechanical hardware.

SPARCS is the first SmallSat to implement dynamic exposure control, in order to avoid detector saturation and preserve the precision of the observations.

The payload will be integrated into BCT's 6U CubeSat and launched to a sun-synchronous terminator orbit for a one-year primary science mission.

SPARCS is a collaboration between the following US institutions and companies: Arizona State University (ASU), Jet Propulsion Laboratory, NASA Goddard Space Flight Center, AZ Space Tech, Blue Canyon Technologies, the University of Arizona at Tucson, the Massachussetts Institute of Technology, Haystack Observatory, Lowell Observatory, Eureka Scientific Inc., and the University of Washington at Seattle.

**Acknowledgements**

The SPARCS project acknowledges the support of the NASA via the APRA program NNH20ZDA001N-APRA. The research was carried out at the Jet Propulsion Laboratory, California Institute of Technology, operated under a contract with NASA (80NM0018D0004). Copyright 2022. All rights reserved..